# Timing is everything: on the stochastic origins of cell-to-cell variability in cancer cell death decisions

Joanna Skommer[1], Subhadip Raychaudhuri[2], Donald Wlodkowic[3]

[1] School of Biological Sciences, University of Auckland, Auckland 1142, New Zealand

[2] Department of Biomedical Engineering, University of California, Davis 95616, USA

[3] Bioelectronics Research Centre, University of Glasgow, Glasgow G12 8LT, UK

Corresponding author:

J.Skommer@auckland.ac.nz, School of Biological Sciences, University of Auckland, Thomas Bld., 3a Symonds Street, Auckland 1142, New Zealand

# 1. Abstract

The diversity of cell populations is regulated by extracellular and intracellular variability. The latter includes genetic, epigenetic and stochastic variability, all contributing to the experimentally observed heterogeneity in response to external death-inducing stimuli. Studies of sources and regulation of variability in commitment to apoptotic cancer cell death are likely to identify the fundamental features of apoptotic protein networks that are responsible for determining the ultimate cell fate. Systems biology approaches, involving computer simulations of the biochemical reactions accompanied, if possible, by experimental verification of selected components of the model, are proving useful in determining the origins of cell-to-cell variability in response to external stress stimuli. Here we summarize our current understanding of the origins of stochastic variability in cells' commitment to apoptosis, and its implications in the field on cancer therapy.

## 2. Introduction

Apoptosis, or programmed cell death, is an evolutionarily conserved mechanism of cell death leading to clean removal of cell corpses due to the exposure of the phagocytosis marker, phosphatidylserine, on the surface of dying cells. Therefore, from the therapeutic point of view, apoptosis is preferred over inflammation-inducing necrotic cell death. Multiple pathways regulate the induction of apoptosis, including the death receptor (extrinsic) pathway, the mitochondrial (intrinsic) pathway, apoptosis induced by DNA damage as well as damage to intracellular organelles other then mitochondria, for example endoplasmic reticulum (1, 2). Our understanding of apoptotic pathways and their key regulators has increased over the last several years, with the key observation that cancer cells have increased expression of one or many of the anti-apoptotic proteins, and/or disrupted function of pro-apoptotic proteins. These observations have led to the development of multiple apoptosis-targeted anti-cancer drug therapies, such as BH3 mimetics designed to block the action of anti-apoptotic Bcl-2 family members, involved in regulation of the intrinsic mitochondrial pathway, or recombinant ligands of death receptors, which induce the extrinsic pathway of apoptosis (3-6). Repeatedly we and others have observed that any of these treatment approaches leads to a highly heterogeneous response in cultured cancer cells, including not only varying kinetics of cell death but also bifurcation in cell fate (some cells undergo cell death, whereas other cells survive), resulting in fractional cell killing. Recent studies by us and others have shed much light on the underlying molecular mechanisms of cell-to-cell variability in the execution of cancer cell death, largely supported by the development of new multiparameter analysis technologies as well as computational approaches (7). Most importantly, it appears that not only genetic and epigenetic differences between cancer cells, but also very subtle changes in protein concentrations and the intrinsic stochastic variability of the reactions within the apoptotic pathway contribute to the observed cell-to-cell variability. Moreover, we are now closer to understanding which parts of the apoptotic pathways introduce most of the variability in cell death decisions.

## 3. Variability during the response to pro-apoptotic stimuli

The response to pro-apoptotic stimuli is affected by the dynamic protein changes after the treatment, as suggested by Cohen and co-authors who analyzed cell-to-cell variability in the expression of about 1000 proteins using CD-tagging approach, which preserves endogenous chromosomal context, in H1299 human lung carcinoma cells (8). The expression of some proteins exhibited a relatively uniform dynamics across the cell population, with the variability attributable to differences in cell-cycle stage and probably due to stochastic processes. The cell-to-cell variability in protein expression increased considerably after administration of the pro-apoptotic drug camptothecin, with several proteins showing bi-modal behavior, i.e. increased expression is some cells, and unchanged or even decreased expression in other cells (8). Interestingly, only few of these proteins were correlated with cell fate. Nevertheless, the study exemplifies how a population of seemingly identical cells can diverge in terms of response to stress stimuli. The mechanisms governing such a divergence require further investigation, with evidence mounting to support the role of stochastic variability in gene expression, and in protein signaling reactions, as discussed below.

**4. Transient heritability**

The regulation of gene expression is based on promoters and transcription factors. The quantitative relation between the amount of active transcription factors and the rate of production (transcription and translation) of gene products fluctuates dynamically in individual cells over a time scale of one cell cycle (9). These fluctuations originate from a composite variability that exists in many cellular components, including random production and degradation of mRNAs and proteins, as well as from the extrinsic noise (9, 10). This indicates that cellular signaling pathways are governed not only by genetic and epigenetic cell-to-cell differences, but also by the inherent variability in protein expression levels originating at the level of gene expression. Compared to genetic or epigenetic differences, the heritability of the qualitative and quantitative state of the proteome is transient. This is also the case for proteins involved in apoptotic signaling. The coefficient of variability (CV) in protein expression is estimated to be in the range of 0.21-0.6, with log-normal distribution in protein levels even in a population of genetically identical cells (11, 12). This variability can explain, at least partially, large cell-to-cell variability in cells' commitment to apoptosis and time-to-death (11). The abundance of cellular components and organelles at the time of division allows their binominal partitioning during cell division (e.g. Fig. 1; 11) (Figure 1– time lapse of mitochondria division between sister cells). Accordingly, sibling cells (which are deemed to be genetically and epigenetically identical) respond nearly synchronously to death stimulus, both during the intrinsic and extrinsic pathway of apoptosis (Figure 2a) (11, 13, 14). Simultaneous response of sibling cells occurs due to transiently inherited protein levels, indicating that differences in protein concentrations contribute largely to cell-to-cell variability in time-to-death (11). Interestingly, using mathematical modeling of the extrinsic pathway of apoptosis, Spencer and co-authors (11) suggest that time-to-death is not determined by variations in the expression of single proteins (specifically death receptor 4 and 5, DISC components, the initiator caspase 8 and 10, or Bcl-2 family member Bid), but by the concerted variation in the concentrations of all these proteins. It remains to be studied whether the variations in the level of multiple proteins, rather then single proteins, correlate with time-to-death, a question that could be answered again by using a mathematical model of apoptotic pathways.

## 5. Stochastic variability in apoptotic reactions

The variability in time-to-death in sibling cells is influenced by the time elapsed since division (11,13) (Figure 2b). Moreover, the increased asynchrony in time-to-death, especially long post division (> 3h), is observed in cells exposed simultaneously to TRAIL and cycloxehimide (inhibitor of protein translation; 11), suggesting that factors other then differences in protein concentrations or states can also contribute to the variability in genetically and epigenetically identical cells. It is possible that the intrinsic 'molecular noise' in apoptotic reactions (15), is responsible for such cell-to-cell variability in time-to-death. This is experimentally very difficult to verify, especially considering that the contribution of the intrinsic stochastic fluctuations may depend on the concentrations of proteins involved in generation of the noise, and that addition of cycloheximide is likely to change sensitivity of cells to apoptotic stimuli. Recent developments in the field of computational systems biology open new door to the modeling of cell-to-cell variability in time-to-death (16). Initial results indicated that inherent stochastic variability in apoptosis signaling reactions is sufficient to generate experimentally observed large cell-to-cell variability and slow cell death (> 10h). The advantage of such *in silico* models is that one can selectively activate one of the two pathways in apoptosis and study its behavior. Results from computational studies demonstrated (17), perhaps for the first time, that large cell-to-cell variability and all-or-none type activation through the intrinsic pathway lead to bi-modal probability distributions in downstream signaling activation. In the Monte Carlo computational study (16), diffusion and reaction of signaling molecules are simulated at an individual molecular level (Figure 3). Such explicit simulation of molecular interactions allowed us to model spatial heterogeneity such as the localization of pro- and anti-apoptotic Bcl-2 like proteins on mitochondrial membranes or formation of multi-molecular apoptosome complexes. Models based on ordinary differential equations (ODEs) cannot simulate such important details of apoptotic signaling. More importantly, ODE-based models cannot capture cell-to-cell stochastic variability in apoptosis signaling that arise solely due to inherent stochastic nature of chemical reactions. Our recent simulations elucidated that over-expression of Bcl-2 protein, as observed in cancer cells, can lead to the generation of only few Bax molecules and stochastic release of cytochrome *c* with large cell-to-cell variability. Interestingly, such stochastic effects in apoptosis are dynamically generated during the time-course of signaling even when large numbers of molecules are initially present in the system. Physiological cellular variations in protein concentrations, which can be easily incorporated in Monte Carlo simulations, induce additional cell-to-cell stochastic variability leading to even slower apoptotic death. However, inherent stochasticity of signaling reactions alone can generate significant cell-to-cell variability with time-scales similar to that observed in recent experiments (18). It thus becomes apparent that even if all intra- and extracellular parameters are kept constant, i.e. creating a hypothetical population of identical cells, the cell death response will still not follow a deterministic pattern, and will occur with different cell-to-cell variability depending on the strength of the stress (Figure 2). This implies that exposure to weak stress stimuli is likely to reveal phenomena not detectable during cell death that is kinetically saturated by strong stress. The effect of intrinsic stochastic variability on apoptosis signaling will be most pronounced at long times (~ 10 hrs) after the induction of an apoptotic stimulus. Moreover, the observation that inherent stochastic variability in the signaling reactions can contribute to variability in time-to-death, particularly when cells are exposed to a limited stress signal, may have serious implications in cancer therapy (17, 18).

**6. Slow induction, fast execution**

One of the inherent variables that can be revealed upon treatment with stress stimuli of different strength is the switch from slow and gradual response (induction phase) to fast all-or-none events of the execution phase. Within the pathway of apoptosis there are steps which are independent of the strength and type of stress stimulus. For example, activation of effector caspases following mitochondrial outer membrane permeabilisation (MOMP) occurs nearly immediately after the onset of MOMP in all affected cells, both during the mitochondrial (intrinsic) and during the death receptor (extrinsic) pathway of apoptosis (19). Caspase activation constitutes an irreversible commitment to cell death. In agreement, using multiparameter flow cytometry employing markers of caspase activity and plasma membrane permeability, we observed that activation of the initiator caspase 9 is detected largely in a population of cells with the loss of plasma membrane integrity, cells which are deemed as already dead (Fig. 4). The existence of such all-or-none mechanism for caspase activation ensures that, at a single cell level, cell death will occur independently on the concentration of signaling molecules. In contrast, signaling upstream of mitochondria appears to contribute significantly to kinetic cell-to-cell variability in apoptosis. The dose-dependent variability in time to MOMP was observed both during the treatment with TRAIL (death receptor ligand), as well as during the treatment with BH3 mimetic that induces the mitochondrial pathway of cell death (18). This behavior of variable delay prior to MOMP, followed by rapid activation of caspases and cleavage of their products, was termed as "variable-delay, snap-action switching" (20). Using computational approach based on ordinary differential equations Albeck and co-authors (20) have shown that the Bax pore formation acts as a form of integrator of stress signals. Using probabilistic computational modeling we can study these processes in even greater detail, for example by analyzing to what extent is the cell-to-cell variability in Bax pore formation preserved post-MOMP (18). Our modeling indicates that even though Bax pore formation generates large cell-to-cell variability in time to death, post-mitochondrial low probability events, such as formation of the apoptosome, may also contribute to the observed variability (16, 18).

Overall, in single cells the gradual response upstream of mitochondria is translated into the binary outcome at the level of caspase activation. The variability within the gradual pre-MOMP response is regulated by the concentrations of proteins (11, 18) as well as by the intrinsic stochastic variability in apoptotic signaling reactions (17).

**7. Clinical implications of cell-to-cell variability in commitment to apoptosis**

The sources of cell-to-cell variability, including genetic, epigenetic, and stochastic, have different mechanisms and time scales, but they all contribute to the experimentally observed heterogeneity in response to pro-apoptotic stimuli. The cancer cell variability in response to chemotherapy may have grave implications. Single cells with a particularly long time-to-death (population outliers) can escape the execution of apoptosis, thus leading to fractional cell killing. It has been shown in several studies that survival of cancer stem cells affects the dynamics of cancer relapse (21). As our appreciation of the abundance of cancer stem cells within the tumor mass increases (22), it becomes apparent that survival of even single cancer stem cell may contribute to the failure of cancer chemotherapy. As a result treatment approaches that do not rely on complete eradication of all tumor cells are being advocated. These approaches include adaptive therapy (23), as well as targeting cell cycle and inducing cancer cell senescence rather then cell death (24). Still, the impact of variability in the proteome, and the intrinsic stochastic variability in the apoptotic reactions, on pharmacology and the treatment of human disease remain unknown. Single-cell data in pharmacological screens combined with computational systems biology approaches are likely to increase our understanding of cell-to-cell variability in response to chemotherapy *in vivo*. Nevertheless, it is still challenging to quantify rapid events at the single cell level using conventional bioanalytical approaches (25, 26). Dissecting stochastic signaling complexity at the single cell level can be greatly improved by employing innovative microfabricated systems (27-30). They provide ways to simultaneously analyze large population of cells whereby the position of every cell is encoded and spatially maintained over extended periods of time, and where each cell can be isolated from others to minimize the influence of extrinsic factors such as cell-to-cell contacts and paracrine signaling (Fig. 5). Microfluidic living-cell arrays that employ the laminar flow under low Reynolds numbers provide an appealing avenue for single-cell positioning into a spatially pre-defined pattern and rapid delivery of reagents to isolated subcellular domains, allowing also for quantitative analysis of the dynamic events at a single cell resolution (29, 30). The potential of single-cell microfabricated static arrays to improve therapeutic screening routines has been reported, but their application in experimental support of current systems biology efforts awaits further exploration (26). We predict that the non-genetic cell-to-cell variability will soon become more recognized as the significant factor contributing to fractional cancer cell killing, and the progress in single-cell analysis of primary patient-derived cells will help to estimate the importance of such variability *in vivo*.

**Figure legends**

**Figure 1** Real-time tracking of the subcellular distribution of mitochondria in dividing cells.

Human $CD34^+$ hematopoietic stem cells were cultured on a specifically designed microfluidic chip where a symmetric stem cell division was analysed using time-resolved imaging. Mitochondria were labelled with TMRM probe (tetramethylrhodamine methyl ester perchlorate; yellow). Note that mitochondria are evenly distributed to sister cells.

**Figure 2** Non-genetic variability in time-to-death.

a) Sibling cells respond synchronously to death stimuli, even though different sets of sister cells respond with large differences in time-to-death. The protein levels are inherited by sibling cells, causing them to undergo cell death simultaneously.
b) The memory of sibling cells is lost as the time post division elapses, leading to decorrelation in time-to-death between sister cells. The longer the time post division, the lower correlation between sibling cells. Treatment with protein synthesis inhibitor slows down decorrelation, but does not stop it.

**Figure 3** Schematics and decision-tree of the Monte Carlo method used in our probabilistic computational simulations of apoptotic signaling.

**Figure 4** Caspase activity is nearly restricted to cells with permeabilized plasma membrane.

Jurkat T cells were treated with a BH3 mimetic HA14-1 (20μM) for 24h, followed by staining with FITC-LEHD-FMK (caspase 9 FLICA) and 7-AAD (marker of plasma membrane permeability) and flow cytometry analysis. Cells with preserved plasma membrane integrity (region R1; 7-$AAD^{neg}$) have nearly no caspase activity detected, whereas cells permeable to 7-AAD (region R2) are labeled with FLICA, indicative of caspase 9 activation. Activation of caspase 9 prior to loss of plasma membrane integrity can only be detected shortly after treatment with a pro-apoptotic stimulus, as activation of caspases constitutes an all-or-none event leading to fast dismantling of the cell.

**Figure 5** Prototypes of cell microarrays for a real-time (4D) analysis of living cells.

a) Design of the static cell microarray that facilitates single cell docking into an array of microfabricated wells. Cells passively sediment and lodge into a pre-defined pattern. Separate cells are inaccessible to neighboring cells which minimizes the influence of extrinsic factors, such as physical cell-to-cell contacts and paracrine signaling. This design facilitates real-time analysis at both single cell and population level.

b) Scanning electron micrograph of the cell microarray fabricated in glass using femtosecond pulse laser $\mu$-machining. Single wells are 25 μm in diameter.

c) Design of the microfluidic cell array that facilitates active (hydrodynamic) cell docking into an array of microfabricated cell traps

d) Scanning electron micrograph of the single cell trap fabricated in a biologically compatible elastomer PDMS. Inset shows a single HL60 cell docked inside a trap and stained with fluorescent probes to detect apoptosis (Annexin V-APC; AV-AP and Propidium Iodide; PI).

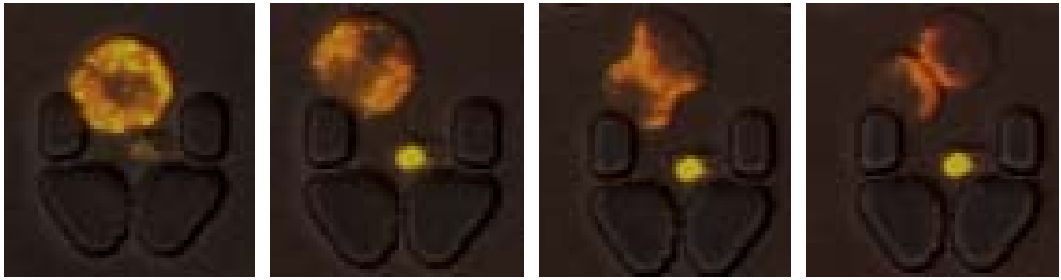

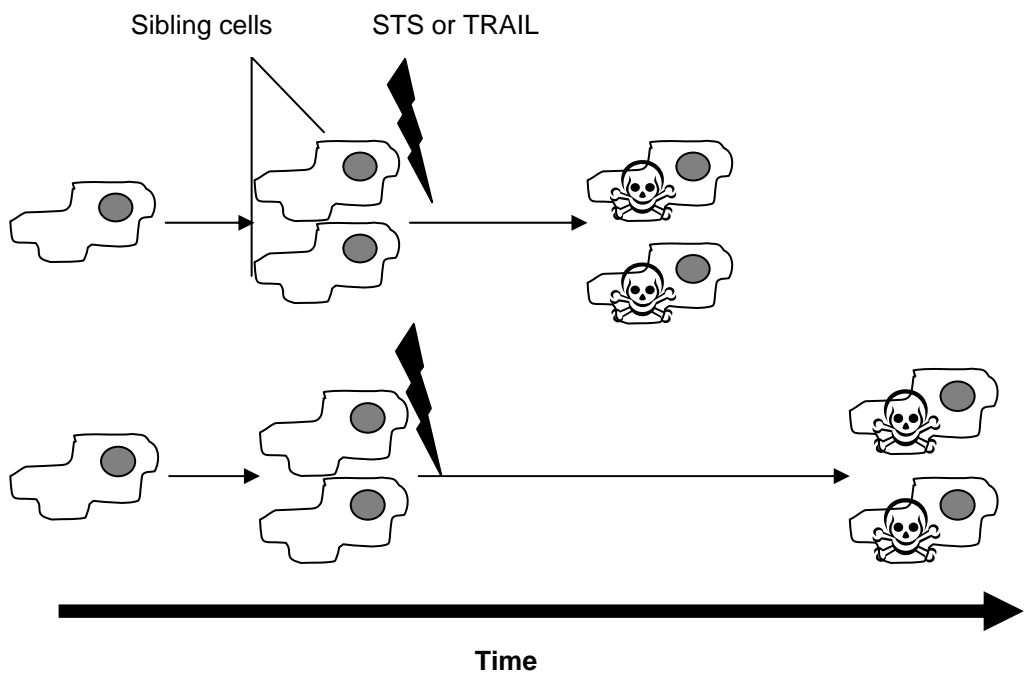

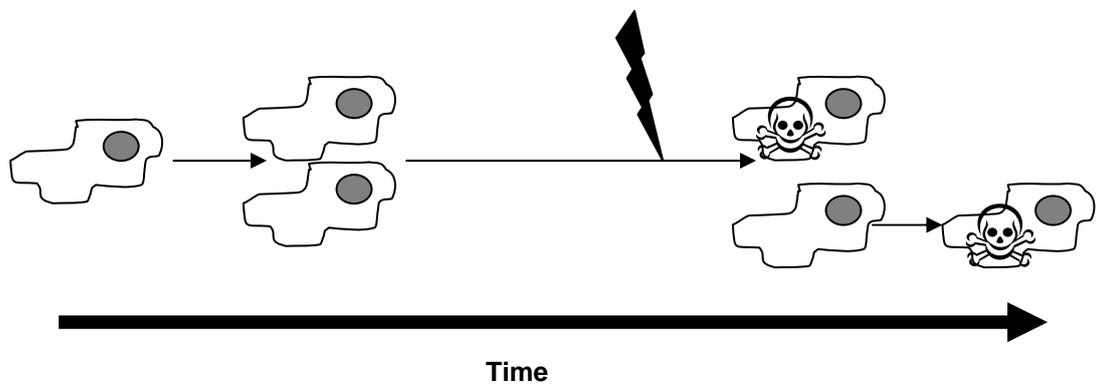

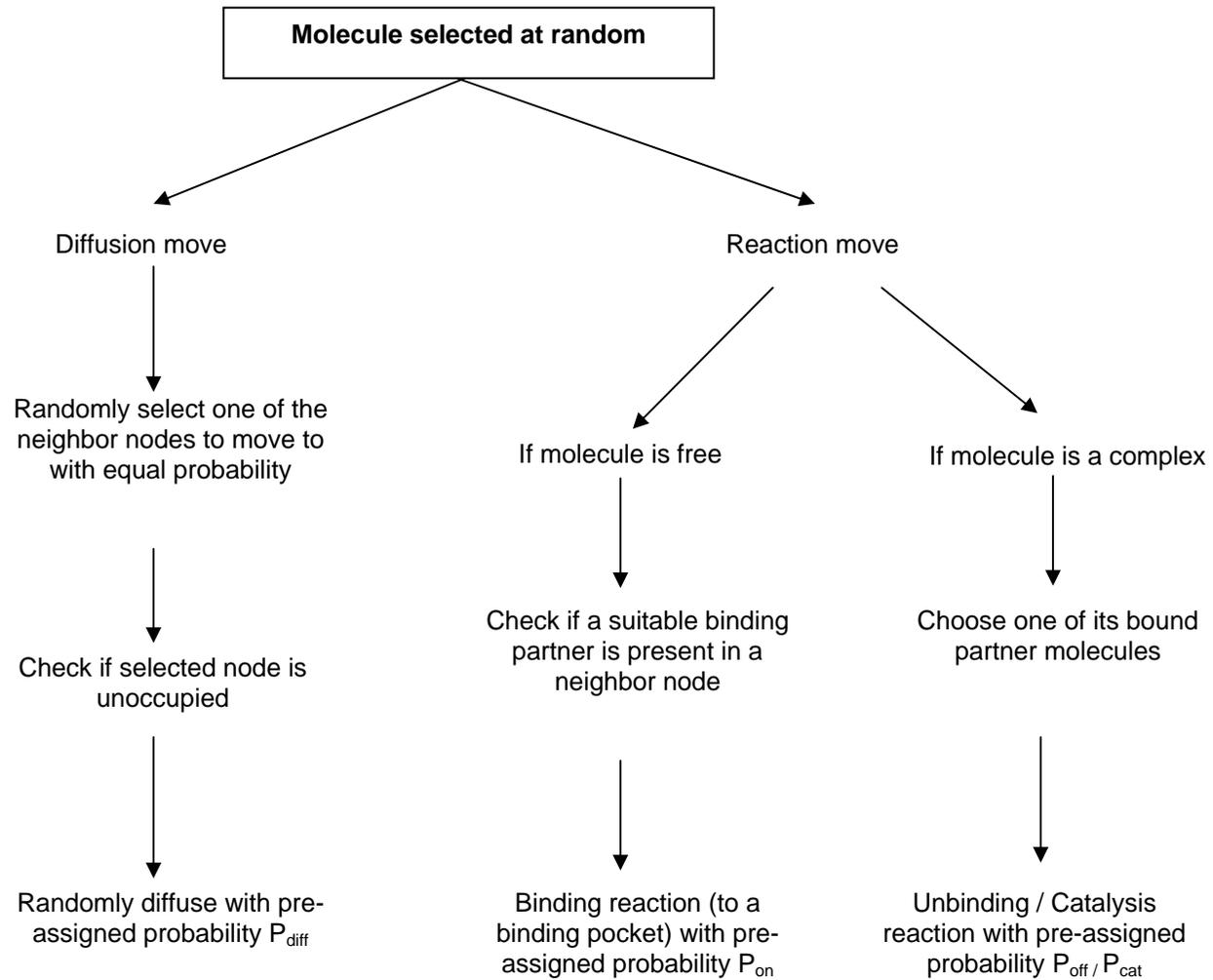

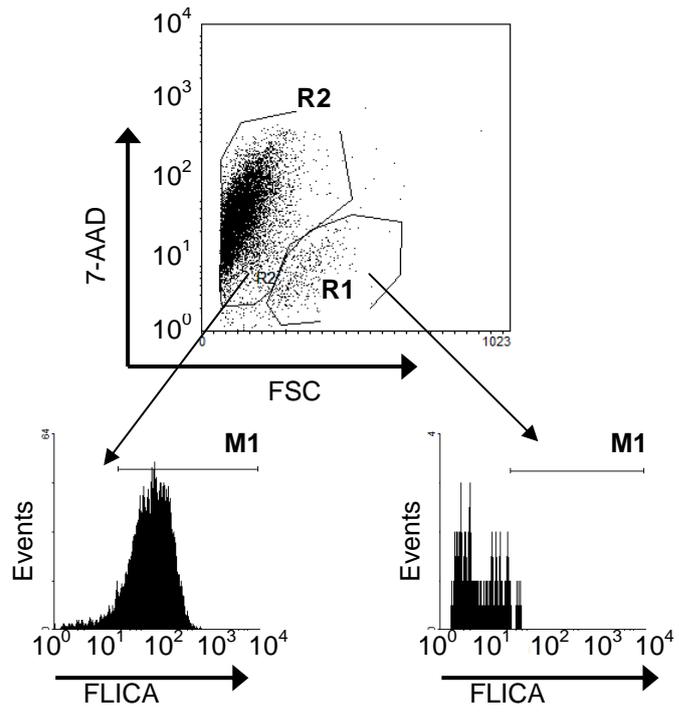

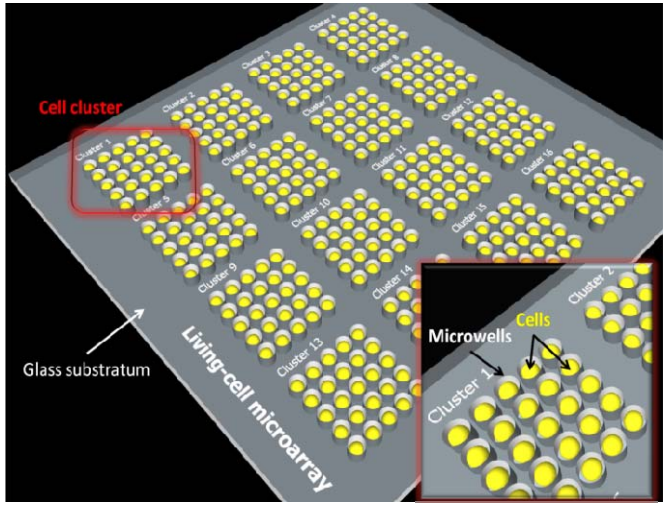

A

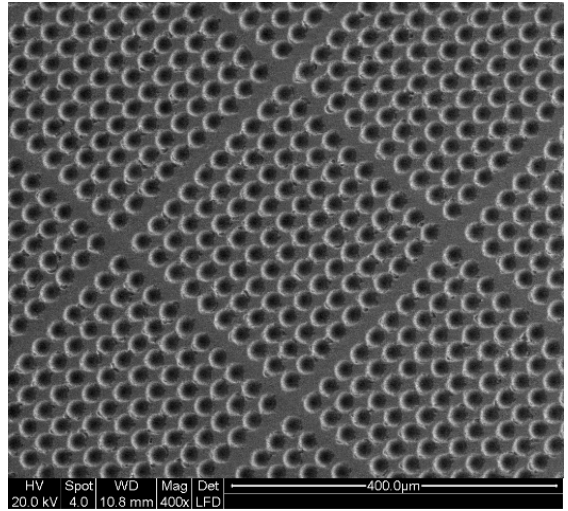

B

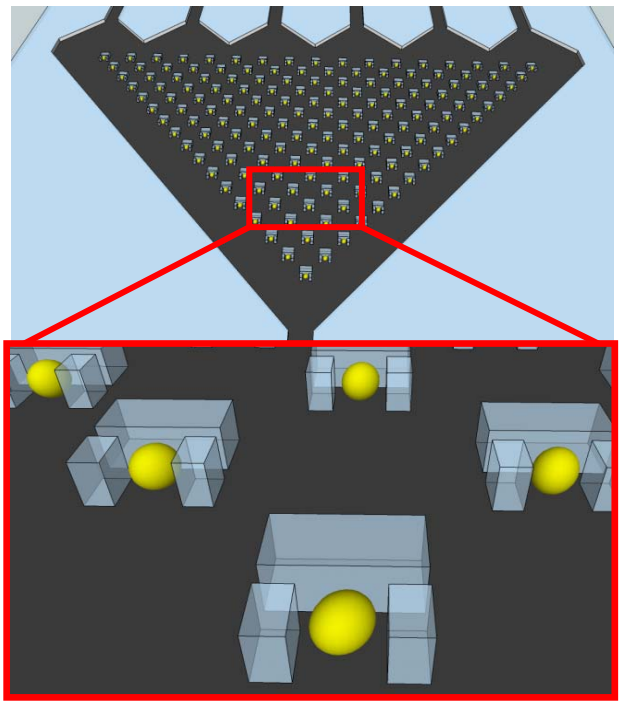

C

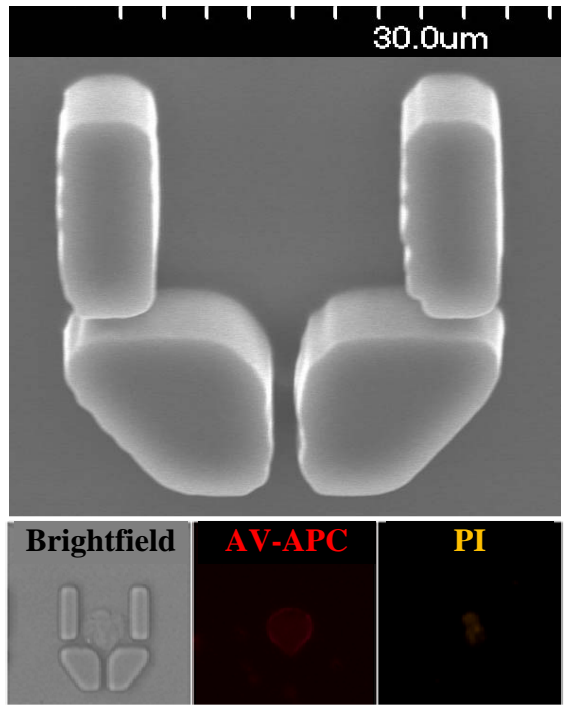

D